\documentclass{aastex701}
\usepackage{amsmath}

\usepackage{booktabs} 
\usepackage{makecell}  
\begin{document}

\title{SETI Observations of k-Hz Periodic Radio Signals from Five Nearby Stars with FAST at L Band}

\correspondingauthor{Tong-Jie Zhang}
\email{tjzhang@bnu.edu.cn}

\author[0009-0002-1805-4288]{Yu Hu}
\email{yuhu_lou@outlook.com}
\altaffiliation{These authors contributed equally to this work.}
\affiliation{Institute for Frontiers in Astronomy and Astrophysics, Beijing Normal University, Beijing 102206, China}
\affiliation{School of Physics and Astronomy, Beijing Normal University, Beijing 100875, China}

\author[0000-0002-8719-3137]{Bo-Lun Huang}
\email{bolunh@hotmail.com}
\altaffiliation{These authors contributed equally to this work.}
\affiliation{Institute for Frontiers in Astronomy and Astrophysics, Beijing Normal University, Beijing 102206, China}
\affiliation{School of Physics and Astronomy, Beijing Normal University, Beijing 100875, China}
\author[0000-0002-8604-106X]{Vishal Gajjar}
\email{vishalg@berkeley.edu}
\affiliation{Breakthrough Listen, University of California, Berkeley, CA 94720, USA}
\affiliation{SETI Institute, 339 N Bernardo Ave Suite 200, Mountain View, CA 94043, USA}

\author[0000-0003-3977-4276]{Xiao-Hang Luan}
\email{202431101067@mail.bnu.edu.cn}
\affiliation{Institute for Frontiers in Astronomy and Astrophysics, Beijing Normal University, Beijing 102206, China}
\affiliation{School of Physics and Astronomy, Beijing Normal University, Beijing 100875, China}

\author[0000-0002-4683-5500]{Zhen-zhao Tao}
\email{tzzzxc@163.com}
\affiliation{College of Computer and Information Engineering, Dezhou University, Dezhou 253023, China}

\author[0000-0002-3363-9965]{Tong-Jie Zhang}
\email{tjzhang@bnu.edu.cn} 
\affiliation{Institute for Frontiers in Astronomy and Astrophysics, Beijing Normal University, Beijing 102206, China}
\affiliation{School of Physics and Astronomy, Beijing Normal University, Beijing 100875, China}

\begin{abstract}
We report a radio SETI search for periodic, kHz-wide signals from five of the nearest stars observable with the Five-hundred-meter Aperture Spherical radio Telescope (FAST). Using the 19-beam L-band receiver (1.05–1.45\,GHz), we obtained 1200\,s tracking observations of Groombridge~34~A/B, Ross~248, 61~Cygni~B, and Ross~128. Dynamic spectra from all beams and both linear polarisations were searched channel by channel with a fast-folding algorithm sensitive to periods between 1.1 and 300\,s. A multi-layer RFI-mitigation pipeline exploits multi-beam occupancy, cross-target bad-channel statistics, XX/YY polarisation coincidence, broad frequency masks, and narrow site-specific RFI exclusion zones, followed by clustering in period–frequency space. The pipeline is validated on FAST observations of PSR~B0329+54, where we recover the known 0.714\,s spin period and harmonic structure in the expected beam. For the stellar sample, successive cuts reduce the raw FFA hit lists ($>10^{6}$ hits per target) to a small number of cluster-level candidates, all of which exhibit clear radio-frequency interference signatures in phase–time and phase–frequency diagnostics. We therefore report no convincing detections of periodic transmitters in our searched parameter space. Using the radiometer equation with our adopted detection threshold (S/N = 25) and assuming a duty cycle $\delta=0.1$, we obtain upper limits of $\simeq(7$–$9)\times10^{9}$\,W on the isotropic-equivalent EIRP of kHz-wide periodic beacons at these stars, among the most stringent constraints to date on periodic radio emission from nearby stellar systems.
\end{abstract}

\section{Introduction}
\label{sec:intro}

The search for extraterrestrial intelligence (SETI) aims to quantify how often technological civilizations arise and persist in the Universe, and to constrain the prevalence of the associated ''technosignatures'' physical manifestations of technology that can be detected over interstellar distances \citep{cocconi1959,tarter2001seti,wright2022techno}.  In the radio band, technosignatures are particularly attractive because engineered transmitters can be both highly luminous and narrowly confined in frequency or time, standing out sharply against natural astrophysical backgrounds.  Since the pioneering suggestion to monitor the 21\,cm hydrogen line for artificial signals \citep{cocconi1959}, most large-scale radio SETI programs have emphasized narrowband, slowly drifting continuous-wave signals in the centimeter band, a strategy now executed at scale by multiple facilities including the Green Bank Telescope (GBT) and FAST \citep[e.g.,][]{huang2025fast}.

Beyond continuous-wave beacons, periodic radio emission is a natural and energetically efficient technosignature morphology.  A rotating or orbiting beacon that produces a narrow pulse once per spin or orbital cycle can deliver high instantaneous flux while maintaining a low time-averaged power budget, analogous to how pulsars are detected despite modest luminosities \citep{ransom2001}.  From the perspective of a distant receiver, such a transmitter would appear as a train of pulses with a well-defined period $P$ and duty cycle $\delta$, potentially confined to a narrow frequency channel but not necessarily dispersion-limited in the same way as natural broadband bursts.  Recent work has demonstrated that a fast folding algorithm (FFA), applied on a per-channel basis to dynamic spectra, is a powerful tool for uncovering such periodic technosignatures.  In particular, the BLIPSS pipeline has carried out the first FFA-based search for periodic alien beacons toward the Galactic Center at 4–8\,GHz, setting upper limits on kHz-wide pulse trains with periods between 11 and 100\,s \citep{suresh2023blipss}.

The Five-hundred-meter Aperture Spherical radio Telescope (FAST) offers an opportunity to extend FFA-based technosignature searches into a new regime of sensitivity and sky coverage.  FAST's 300\,m illuminated aperture, coupled to a 19-beam L-band receiver covering roughly 1.05–1.45\,GHz, provides system temperatures of $\sim$25\,K and an effective sensitivity that surpasses that of previous single-dish facilities at similar frequencies \citep{jiang2020fastperf}. The multi-beam receiver architecture enables powerful radio-frequency interference (RFI) rejection based on spatial coincidence patterns across beams, while the high spectral resolution and stability of the L-band backend support coherent searches for faint, slowly repeating signals.  FAST has already been used for narrowband technosignature searches, including the dedicated exoplanet SETI observations with the 19-beam receiver \citep{tao2022sensitive} and more, but its potential for periodic technosignature searches with FFA-based methods has not yet been fully explored.

\begin{figure}
    \centering
    \includegraphics[width=0.7\linewidth]{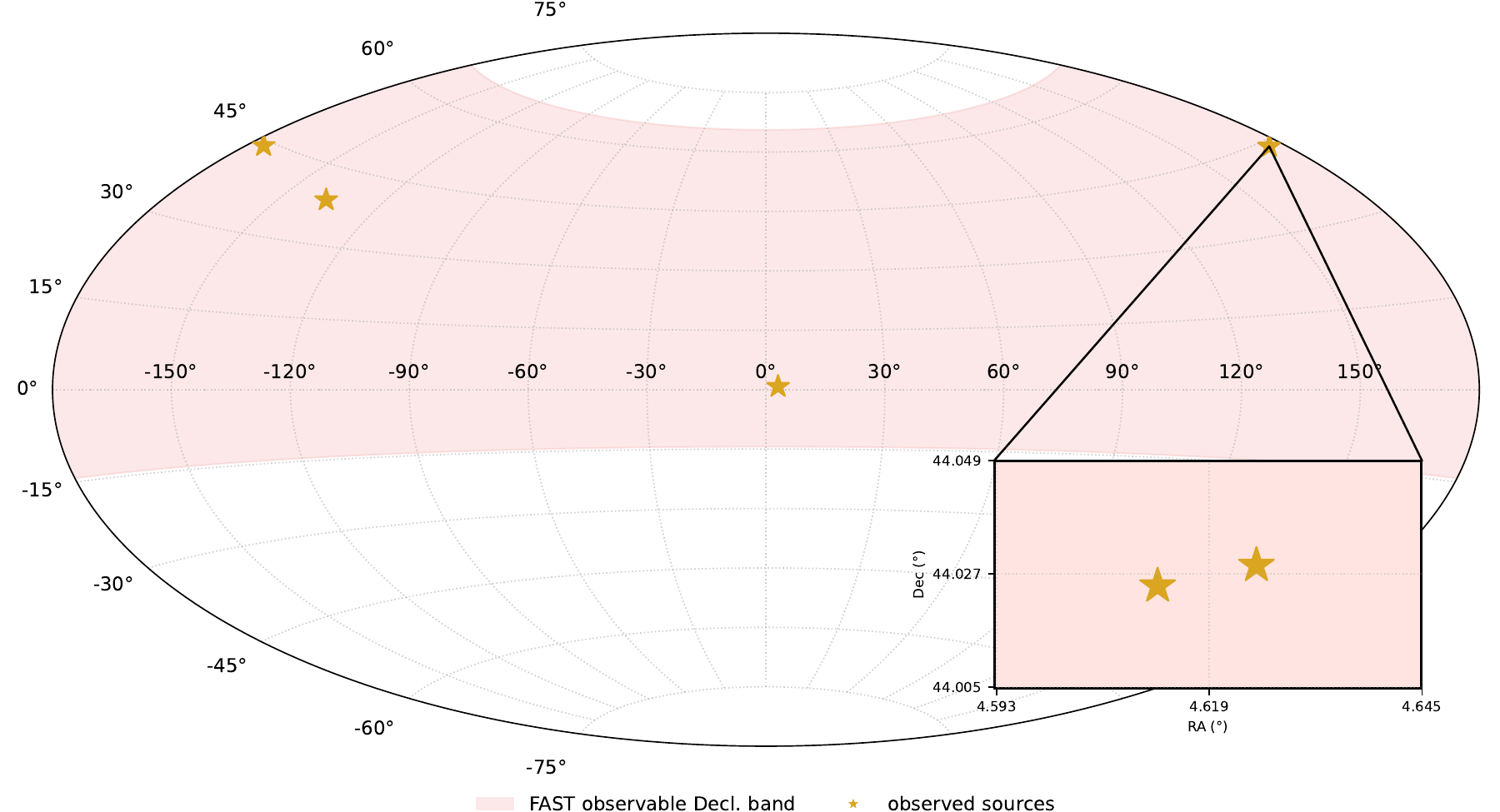}
    \caption{The positions of the five target stars on the FAST sky coverage map}
    \label{fig:pos}
\end{figure}

In this work, we present a search for periodic technosignatures toward five of the nearest stellar systems (as shown in Fig.\ref{fig:pos}) observable with FAST: Groombridge~34~A, Groombridge~34~B, Ross~248, 61~Cygni~B, and Ross~128.  All five stars lie within $\sim 3$–4\,pc, are low-mass M or K dwarfs with long main-sequence lifetimes, and have either confirmed or strongly suspected low-mass planets, making them long-standing benchmarks for studies of nearby potentially habitable systems \citep[e.g.,][]{bonfils2018ross128b,turnbull2003habcat}.  Their proximity means that even modestly powered transmitters or Earth-level radio leakage would be detectable with FAST: for isotropic 1\,kHz-wide beacons at L-band, our non-detections translate into equivalent isotropic radiated power (EIRP) limits of order $10^{9}$\,W, which is few orders of magnitude below the most powerful human-made transmitters.  The detailed stellar and planetary properties, as well as the observing strategy and scheduling of the 19-beam tracks, are summarized in Section~\ref{sec:obs}.

Methodologically, our analysis builds upon and extends the BLIPSS framework \citep{suresh2023blipss} to the case of FAST multi-beam L-band data.  We apply a per-channel FFA search over periods from 1.1 to 300\,s, using dual linear polarizations and all 19 beams, and then pass the resulting hits through a multi-stage RFI discrimination and clustering pipeline. To validate the end-to-end performance of this pipeline in a regime of known astrophysical periodicity, we analyze FAST observations of the bright pulsar B0329+54 with the same BLIPSS configuration.
The remainder of this paper is organized as follows. In Section~\ref{sec:obs} we describe the FAST observations, instrumental setup, and the selection of the five nearby targets. Section~\ref{sec:analysis} presents the data conditioning, FFA-based periodic-search pipeline, and multi-layer RFI excision strategy, and includes an end-to-end validation on PSR~B0329+54. In Section~\ref{sec:discussion} we report the results of the search, quantify our sensitivity and EIRP limits, and place these constraints in the context of previous radio SETI efforts. We summarize our conclusions and outline prospects for future FAST periodic SETI surveys in Section~\ref{sec:conclusion}.

\section{Observations and Targets}
\label{sec:obs}

\subsection{FAST and the L-band multibeam system}

All observations were carried out with the Five-hundred-meter Aperture Spherical radio Telescope (FAST) in Guizhou, China \citep{jiang2020fastperf}. FAST operates with an illuminated aperture of $\sim 300$~m at L band, delivering a gain of $G \approx 16$~K~Jy$^{-1}$ and a system temperature $T_{\rm sys} \simeq 25$~K for the central beam of the 19-beam receiver \citep{jiang2020fastperf,zhang2020fastseti}. The corresponding system-equivalent flux density is ${\rm SEFD} \sim 2$~Jy, providing exceptional sensitivity to narrowband or quasi-periodic signals from nearby stars.

We used the cryogenic L-band 19-beam receiver, which provides simultaneous coverage of 19 beams in the focal plane: one central beam (M01) and 18 surrounding beams (M02--M19). At a representative frequency of 1.25~GHz, the half-power beam width of each beam is $\sim 2\farcm9$ \citep{jiang2020fastperf}. The central beam was pointed at each target star; the outer beams were used as OFF-source references for radio-frequency interference (RFI) rejection (Section~\ref{sec:analysis}). 

The receiver was configured to cover the nominal FAST L-band frequency range of 1.05--1.45~GHz \citep{jiang2020fastperf}. The backend was operated in search mode using the spectral processor SPEC(F), producing 1,024k ($1{,}048{,}576$) frequency channels across the 400~MHz instantaneous bandwidth. This corresponds to a frequency resolution of 476.84~Hz per channel. Two orthogonal linear polarizations (XX and YY) were recorded independently for each beam. The integration time per spectrum was 0.1~s.

Raw data were recorded in FITS format and subsequently converted to filterbank files, generating one filterbank per beam and per polarization. All subsequent analysis in this work (conditioning, Fast Folding Algorithm search, and multi-layer RFI excision) was performed on these filterbank files using the BLIPSS software framework \citep{suresh2023blipss}, as detailed in Section~\ref{sec:analysis}.

\subsection{Target selection and stellar properties}
\label{sec:targets}

Our stellar sample consists of five nearby low-mass stars within $\sim 4$~pc: Groombridge~34~A, Groombridge~34~B, Ross~248, 61~Cygni~B, and Ross~128. Rather than attempting to construct a statistically complete catalog, we adopt a pragmatic selection designed to maximize the scientific leverage of a FAST periodic technosignature search. Targets are required to satisfy the following criteria: (1) distances of a few parsecs, so that FAST can place stringent limits on the equivalent isotropic radiated power (EIRP) of any putative transmitters; (2) spectral types K--M, which dominate the local stellar population and are natural hosts for long-lived habitable environments (\cite{Valle_2014} \cite{Kopparapu_2013}).(3) good visibility from FAST with high elevation angles, given its declination limits of approximately $-14^{\circ} \lesssim \delta \lesssim +66^{\circ}$ \cite{qian2020fastreview}; and (4) existing evidence for either confirmed low-mass planets or strong exoplanet/habitability interest in the nearby-star literature (\cite{turnbull2003habcat,bonfils2018ross128b}).

Within these criteria, we chose a diverse but astrophysically motivated mix of systems. The sample includes: a nearby early M-dwarf multi-planet system (Groombridge~34~A) with well-studied radial-velocity companions \citep{howard2014gj15ab,pinamonti2018gj15a}; its wide M-dwarf companion (Groombridge~34~B), which lacks confirmed planets and thus serves as a close control in essentially the same local environment; a very low-mass, planet-free nearby M dwarf (Ross~248), representative of late-type red dwarfs targeted in many SETI and exoplanet surveys \citep{gautier2007ross248,turnbull2003habcat}; a nearby K dwarf (61~Cygni~B), chosen as a representative ''Goldilocks'' host star with a long-lived habitable zone and extensive observational history \citep{turnbull2003habcat}; and a quiet M dwarf hosting a temperate Earth-mass planet (Ross~128~b) near the inner edge of the classical habitable zone \citep{bonfils2018ross128b}. This combination is not intended to be statistically complete, but rather to span a range of stellar types and planetary architectures that are frequently discussed in the context of habitable environments and technosignatures.

Accurate astrometry and distances for all five targets are taken from \emph{Gaia}~DR3 \citep{gaia2023dr3}. The \emph{Gaia} identifiers, equatorial coordinates, and adopted distances are listed in Table~\ref{tab:targets}.

\begin{deluxetable*}{llcccccc}
\tablecaption{FAST targets and observing logistics\label{tab:targets}}
\tablehead{
\colhead{Target} &
\colhead{Other name(s)} &
\colhead{Gaia DR3 ID} &
\colhead{SpT} &
\colhead{R.A. (J2000)} &
\colhead{Dec. (J2000)} &
\colhead{$d$ (pc)} &
\colhead{Obs.\ window (UTC, 2023-11-23)}
}
\startdata
Groombridge~34 A &
GX~And, GJ~15A &
385334230892516480 &
M1.5~V &
00:18:27.04 &
+44:01:28.9 &
3.5625 &
23:04:00--23:24:00 \\
Groombridge~34 B &
GQ~And, GJ~15B &
385334196532776576 &
M3.5~V &
00:18:29.94 &
+44:01:43.3 &
3.5638 &
23:29:00--23:49:00 \\
Ross~248 &
HH~And, GJ~905 &
1926461164913660160 &
M6~V &
23:41:55.20 &
+44:10:14.1 &
3.4947 &
22:32:00--22:52:00 \\
61~Cyg~B &
HD~201092, GJ~820B &
1872046574983497216 &
K7~V &
21:07:00.72 &
+38:45:20.1 &
3.1598 &
20:05:00--20:25:00 \\
Ross~128 &
FI~Vir, GJ~447 &
3796072592206250624 &
M4~V &
11:47:45.12 &
+00:47:57.4 &
3.3731 &
08:55:00--09:15:00 \\
\enddata
\tablecomments{
Spectral types and distances follow standard nearby-star compilations and Gaia~DR3-based catalogs, rounded to match the precision adopted in the analysis. The observing windows list the UTC tracking intervals for each 1200\,s FAST pointing.
}
\end{deluxetable*}

\subsection{Observing setup and schedule}

All five stellar targets were observed on 2023~November~23 in a pointed tracking mode. For each star we obtained a single 1200~s (20~min) integration with the 19-beam L-band receiver, keeping the central beam M01 locked on the target position. The outer beams sampled adjacent sky regions at separations of a few arcminutes, providing a multi-beam view of the local RFI environment. The start and end times for each observation, together with the adopted coordinates and distances, are listed in Table~\ref{tab:targets}. 

For each target, the telescope elevation was kept as high as practical within FAST operational constraints, taking advantage of the accessible declination range of $-14.6^{\circ}$ to $+65.6^{\circ}$ \citep{qian2020fastreview}. This strategy minimizes atmospheric opacity and ground spillover while maximizing gain stability \citep{jiang2020fastperf}. The front-end gain and system temperature were monitored by the standard FAST calibration procedures; these values were later propagated into our EIRP constraints (Section~\ref{sec:discussion}).

The backend configuration was identical for all targets and for the validation pulsar described below: dual polarizations, 1.05--1.45~GHz coverage, 1,024k channels, and 0.1~s spectral integration time. The data stream from each beam and polarization was written to a separate PSRFITS file, resulting in 38 filterbank files per target (19 beams $\times$ 2 polarizations).

\subsection{Validation dataset: PSR B0329+54}
\label{sec:psr_obs}

In addition to the five stellar targets, we observed the bright pulsar PSR~B0329+54 as an internal validation dataset. PSR~B0329+54 is one of the brightest Northern Hemisphere radio pulsars, with a spin period $P \approx 0.714$~s and well-determined timing ephemerides from the ATNF pulsar catalogue \citep{manchester2005atnf}. Its strong, stable radio pulses make it an ideal test source for validating periodicity searches in the time and frequency domains.

We acquired a 1200~s pointed track on PSR~B0329+54 using the same receiver, bandwidth, channelization, polarization setup, and integration time as for the stellar targets. The pointing placed the pulsar at the center of beam M01, with the outer beams sampling nearby sky. The resulting filterbank files were processed through the identical BLIPSS-based pipeline, including the time-series conditioning, Fast Folding Algorithm search, and the multi-layer RFI and consistency filters described in Section~\ref{sec:analysis}. 

By comparing the recovered period, harmonic structure, and frequency-dependent S/N patterns of PSR~B0329+54 against expectations from high-precision timing solutions, we verify both the instrumental configuration and the internal consistency of our analysis pipeline. This validation step is essential to demonstrate that any non-detections in the stellar sample reflect astrophysical reality rather than shortcomings in our search methodology.

\section{Data Conditioning, Periodic-search Pipeline, and RFI Excision}
\label{sec:analysis}

\begin{figure}
    \centering
    \includegraphics[width=0.8\linewidth]{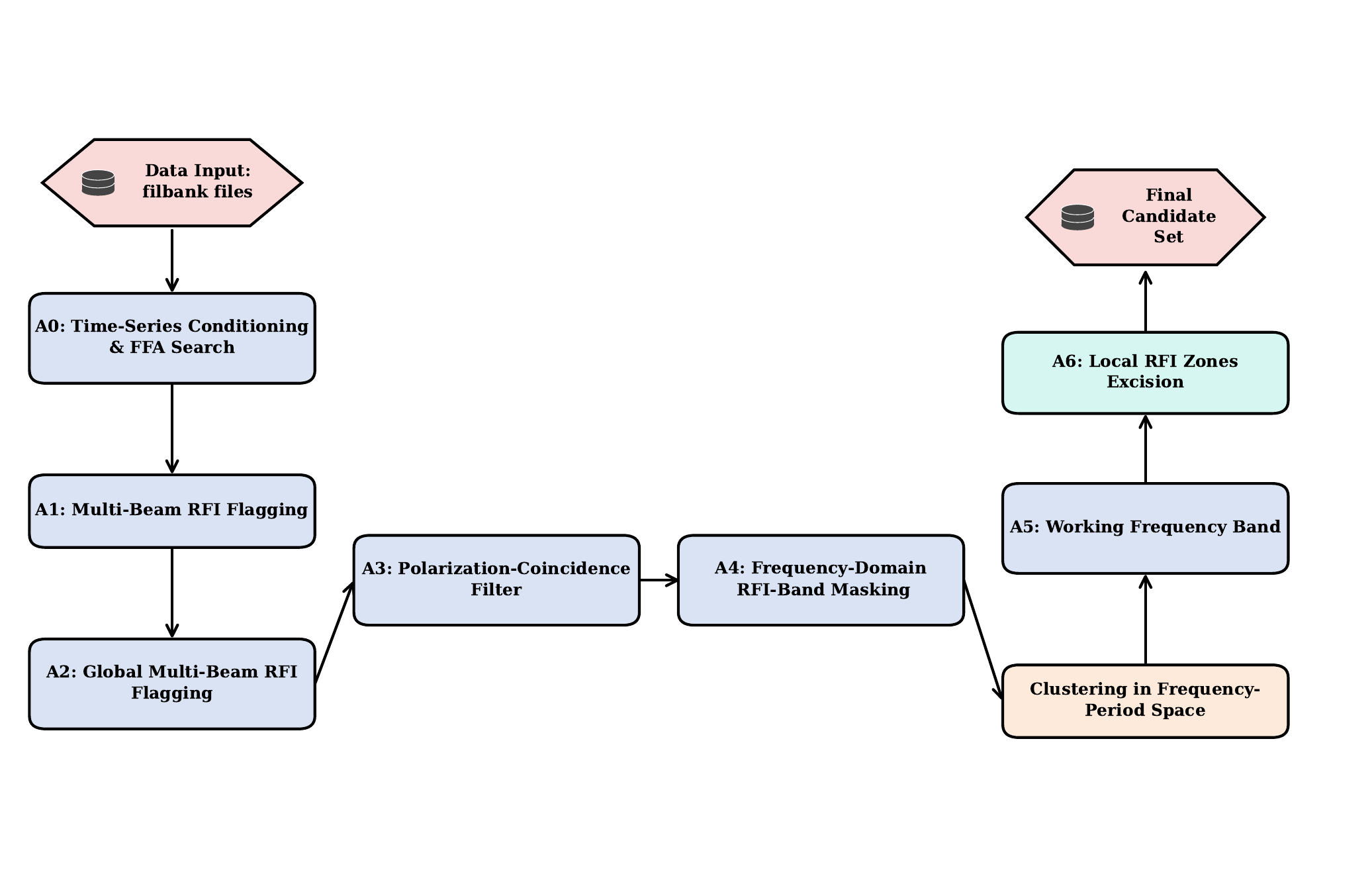}
\caption{\textbf{Schematic workflow of the data analysis and RFI mitigation pipeline.} 
The sequence progresses from the initial search (Stage A0) through multi-beam (A1), cross-target (A2), polarization (A3), and frequency-domain (A4) masking, followed by the working-band excision (A5). 
The ``A6 Local RFI Zones Excision'' step is an additional refinement applied exclusively to the 61 Cyg B dataset to remove persistent target-specific interference (see Section \ref{sec:rfizones}). 
The hit count reductions resulting from each filtering stage (A0--A5) are detailed in Table \ref{tab:rfi_flagging_aastex}.}
\label{fig:pipeline_flowchart}

    \label{fig:RFI}
\end{figure}

Our analysis pipeline starts from the filterbank files described in Section~\ref{sec:obs}. For each target, beam, and polarization, we construct a time series at each frequency channel, apply basic conditioning and detrending, and then perform a per-channel Fast Folding Algorithm (FFA) search for strictly periodic signals using the BLIPSS software package \citep{suresh2023blipss}. The FFA stage produces a large set of candidate periods and associated signal-to-noise ratios (S/N) as a function of frequency. Figure\ref{fig:RFI} shows the workflow for RFI flagging techniques we have applied.

\subsection{Time-series conditioning and FFA search}
\label{sec:analysis_ffa}

For each filterbank file, we perform an FFA search in each channel using BLIPSS with the configuration optimized for technosignature searches toward nearby stars. The search parameters are given in Table \ref{tab:ffa_params}.
\begin{table}[htbp]
\centering
\caption{FFA search parameter settings used in this work.}
\label{tab:ffa_params}
\begin{tabular}{ll}
\hline
Parameter & Value / Description \\
\hline
Minimum trial period, $P_{\min}$ 
& $1.1\,\mathrm{s}$ \\

Maximum trial period, $P_{\max}$ 
& $300\,\mathrm{s}$ \\

Minimum number of cycles, $f_{p,\min}$ 
& $4$ (within $T_{\rm obs}=1200\,\mathrm{s}$; $P \leq T_{\rm obs}/f_{p,\min}$) \\

Number of phase bins 
& $N_{\rm bin} = 10$--$11$ \\

Duty-cycle resolution 
& $\Delta\delta \simeq 0.1$ \\

Maximum duty cycle 
& $\delta_{\max} = 0.5$ \\

Detection threshold 
& $\mathrm{S/N}_{\rm thr} = 25$ \\
\hline
\end{tabular}
\end{table}

This configuration adapts the successful BLIPSS search toward the Galactic Center \citep{suresh2023blipss}, but is further optimised to the shorter integration time and different period range of our FAST observations. We assume that any technosignature from a $\sim$3--4\,pc system would exhibit negligible dispersion at 1--1.5\,GHz and therefore do not apply any dedispersion; the search is effectively conducted at dispersion measure $\mathrm{DM} \approx 0$.

For each beam and polarization, the FFA search yields a list of hits characterized by the channel index (and hence frequency), trial period, peak S/N, associated pulse width, and several quality metrics. The union of these hit lists over all beams and both polarisations for a given target constitutes the raw hit set, which we designate as \textbf{A0}. This A0 set is the starting point for all subsequent RFI-mitigation stages.

\subsection{Multi-beam RFI flags and cross-target contaminants}
\label{sec:rfi_primary}

Because FAST observes all 19 beams simultaneously, any narrowband or periodic signal that appears in beams that are separated by sufficient angular distance at the same frequency and period is overwhelmingly likely to be of terrestrial origin \citep{luan2023multibeam}. We therefore define a multi-beam occupancy \emph{Code} and associated RFI flags for each candidate hit. In the FAST 19-beam receiver, we follow the strategy of \citet{tao2022sensitive} and select the central beam as the ''ON'' position and a subset of outer beams as ''OFF'' positions for RFI discrimination as \cite{Sheikh_2021} did.

For a given target and frequency--period pair $(\nu,P)$, we examine the FFA hit lists from all beams for a given polarization and mark a beam as ''1'' if a hit is present above the S/N threshold within the matching tolerances described in Section~\ref{sec:pol_filter}, and ''0'' otherwise. This yields a 7-bit binary string(since we only need the centre beam and the six outermost beams), the \emph{Code}, whose $i$-th bit encodes the occupancy of beam M0$i$. A Code of $1000000$ indicates that only the central ''ON'' beam M01 reports a detection, while Codes with multiple ''1'' bits correspond to signals detected in several beams.

We use the Code to define the primary multi-beam RFI flag, $\mathrm{RFI1}$, as
\begin{equation}
    \mathrm{RFI1} = 
    \begin{cases}
        1, & \text{if the number of ''1'' bits in Code} \ge 2, \\
        0, & \text{otherwise.}
    \end{cases}
\end{equation}
Hits with $\mathrm{RFI1} = 1$ are considered to be local RFI and are used only to characterize the RFI environment; they are excluded from the astrophysical candidate stream.

Applying the $\mathrm{RFI1}$ mask to A0 defines the first filtered hit set, \textbf{A1}, consisting of all hits with $\mathrm{RFI1} = 0$.

To identify channels that are persistently contaminated across multiple targets, we define a secondary flag, $\mathrm{RFI2}$, based on cross-target statistics. Because all five stellar observations were obtained within a single day with an identical instrumental setup, we aggregate the frequency channels of all multi-beam RFI across the sample: any hit in any target that is co-located in frequency with a channel where $\mathrm{RFI1}=1$ in \emph{any} observation is marked as
\begin{equation}
    \mathrm{RFI2} = 1.
\end{equation}
The remaining hits have $\mathrm{RFI2} = 0$. This procedure identifies a set of long-term ''bad'' channels, typically associated with fixed-frequency terrestrial or satellite transmissions. Applying the $\mathrm{RFI2}$ mask to A1 yields \textbf{A2}, the subset of hits with $\mathrm{RFI1}=\mathrm{RFI2}=0$.

\subsection{Polarization-coincidence filter}
\label{sec:pol_filter}

A genuine astrophysical signal is expected to be present, though possibly with different amplitudes, in both orthogonal linear polarizations. In contrast, many instrumental artefacts and some classes of RFI appear preferentially in only one recorded polarization. We therefore implement a polarization-coincidence filter introduced by \cite{Tao_2023} and \cite{luan2023multibeam} to distinguish candidates that are consistent between XX and YY from single-polarization hits.

Starting from the per-beam hit lists for XX and YY, we construct a combined list by matching hits in the two polarizations in the $(\nu,P)$ plane. Two hits are considered a match if their frequencies and periods satisfy
\begin{align}
    |\Delta \nu| &\le 0.001~\mathrm{MHz}, \\
    |\Delta P| &\le \max\bigl(0.1~\mathrm{s},\, 0.01\,P\bigr),
\end{align}
where $P$ is the candidate period. We perform a two-way nearest-neighbour match, XX$\rightarrow$YY and YY$\rightarrow$XX, to ensure symmetry and to avoid spurious assignments in crowded regions of parameter space.

For hits that find a counterpart in the other polarization within these tolerances, we set
\begin{equation}
    \mathrm{RFI3} = 0,
\end{equation}
and retain their combined properties (e.g.\ average frequency, period, and S/N). Hits that do not find a counterpart are marked as
\begin{equation}
    \mathrm{RFI3} = 1.
\end{equation}
We consider $\mathrm{RFI3} = 0$ hits as astrophysically plausible in terms of polarization consistency, while $\mathrm{RFI3}=1$ hits are treated as likely instrumental or RFI artefacts. At this stage we also assign a qualitative ''color'' label to each hit for visualization (e.g.\ green for $\mathrm{RFI1}=\mathrm{RFI2}=\mathrm{RFI3}=0$, magenta for $\mathrm{RFI3}=1$, etc.), but these colors are not used in any quantitative selection criteria.

Applying the polarization-coincidence filter to A2 and retaining only hits with $\mathrm{RFI3}=0$ defines the hit set \textbf{A3}, i.e.\ the subset of A2 for which
\begin{equation}
    \mathrm{RFI1} = 0,\quad \mathrm{RFI2} = 0,\quad \mathrm{RFI3} = 0.
\end{equation}
These hits form the basis of our subsequent frequency-domain masking and clustering.

\subsection{Frequency-domain RFI-band mask}
\label{sec:rfiband}

The combination of $\mathrm{RFI1}$ and $\mathrm{RFI2}$ identifies channels and frequency ranges that are frequently contaminated, but does not yet account for the finite width of real-world RFI bands. To construct a conservative frequency-domain mask, $\mathrm{RFI\_band}$, we proceed in three steps.

First, we identify all channels that are ever flagged as $\mathrm{RFI1}=1$ or $\mathrm{RFI2}=1$ in any target and polarization. These channels are sorted by index and adjacent channels are merged into continuous ''core'' RFI segments. Second, for each core segment we expand its boundaries by an integer number $\Delta$ of channels on each side to account for spectral leakage and band-edge effects. Third, we determine an optimal value of $\Delta$ by examining how the surviving fraction of ''good'' hits depends on $\Delta$.

We define as ''good'' those hits that satisfy
\begin{equation}
    \mathrm{RFI1} = 0,\quad \mathrm{RFI2} = 0,\quad \mathrm{RFI3} = 0.
\end{equation}
For a trial expansion $\Delta$, we then compute the fraction of good hits that lie outside all expanded RFI segments. As $\Delta$ increases from 0 upward, this surviving fraction decreases. We adopt as our final expansion width the smallest $\Delta$ for which the marginal loss of good hits per increment in $\Delta$ falls below a chosen threshold (typically $\sim 5\%$ per step). This procedure balances aggressive RFI removal against the preservation of as much clean bandwidth as possible.

Channels lying within the expanded segments are assigned
\begin{equation}
    \mathrm{RFI\_band} = 1,
\end{equation}
and are excluded from subsequent astrophysical candidate selection; all other channels have $\mathrm{RFI\_band} = 0$. Applying this mask to A3 defines the hit set \textbf{A4}, consisting of all hits that satisfy
\begin{equation}
    \mathrm{RFI1} = 0,\quad \mathrm{RFI2} = 0,\quad \mathrm{RFI3} = 0,\quad \mathrm{RFI\_band} = 0.
\end{equation}
The \emph{RFI\_band} mask typically removes a modest fraction of the nominal 1.05--1.45~GHz band, concentrating on a small number of broad RFI complexes.

\subsection{Clustering in frequency--period space}
\label{sec:clustering}

Even after the application of multi-beam, cross-target, polarization, and frequency masks (i.e.\ within A4), the FFA search yields many redundant hits that correspond to the same underlying periodic signal detected at nearby trial periods or frequencies. To avoid double-counting and to simplify the interpretation of non-detections, we compress the surviving hits into clusters in the $(\nu,P)$ plane.

We group the A4 hits by channel index (i.e.\ by frequency). Within each channel, we sort the hits by period and perform one-dimensional clustering in $P$. Two adjacent hits in the sorted list are assigned to the same cluster if their periods satisfy
\begin{equation}
    |\Delta P| \le \max\bigl(0.1~\mathrm{s},\, 0.001\,P_{\rm ref}\bigr),
\end{equation}
where $P_{\rm ref}$ is the period of the earlier hit in the pair. If this condition is not met, a new cluster is started. This scheme captures both nearly identical period estimates and a modest range of harmonically related or aliased periods.

For each cluster, we select the hit with the highest S/N as the cluster representative and record the number of members, the minimum and maximum periods within the cluster, and the counts of contributing XX and YY hits. The result is a cluster-level catalog in which each entry represents a distinct candidate periodicity at a given frequency. We retain both the per-hit (A4) and per-cluster catalogs for downstream checks, but use the cluster-level catalog---restricted to the working frequency band and single-beam detections, as described below---as the basis for our final candidate selection and EIRP limits.

\subsection{Working frequency band and single-beam selection}
\label{sec:working_band}

In practice, certain edge regions of the FAST's 1.00--1.50~GHz band (1.00--1.05~GHz and 1.45--1.50~GHz)are severely contaminated by residual RFI and instrumental systematics than others, even after application of the $\mathrm{RFI\_band}$ mask and are usually excluded for research. There is a ''working band'' within 1050--1450~MHz that balances spectral cleanliness against total bandwidth.

Within this working band, we further restrict attention to strictly single-beam detections as indicated by the multi-beam occupancy Code. Specifically, we require
\begin{equation}
    \mathrm{Code} = 1000000,
\end{equation}
i.e.\ only the on-axis beam M01 reports a detection at the candidate $(\nu,P)$, and all outer beams are consistent with noise. This cut is designed to eliminate signals that might be entering through the far sidelobes or arising from local RFI that illuminates multiple beams.

Applying these criteria to the A4 hits yields the hit set \textbf{A5}: the subset of A4 that lies within 1050--1450\,MHz and satisfies the single-beam condition. The cluster-level catalog constructed from these A5 hits constitutes our ''working-band candidate set''. The hit counts A0--A5 for each target are summarized in Table~\ref{tab:rfi_flagging_aastex}; Section~\ref{sec:result} discusses the corresponding reduction factors and their implications.

\subsection{Local RFI zones and final candidate set}
\label{sec:rfizones}

Although the $\mathrm{RFI\_band}$ mask effectively removes the broad complexes of persistent RFI, the working band still contains a small number of narrow, strongly contaminated frequency ranges. These typically correspond to fixed-frequency emitters that are not sufficiently widespread in channel space or across targets to trigger $\mathrm{RFI2}$, but nevertheless dominate the local hit statistics.

We identify such local RFI zones by inspecting the distribution of working-band candidates in the frequency–hit-count plane, using robust measures (e.g.\ median absolute deviation) to flag channels whose hit counts greatly exceed the local median. Adjacent contaminated channels are merged into narrow exclusion intervals, typically spanning $\lesssim 10$~MHz each. Based on this procedure, we define a small set of frequency ranges which we excise from the working band; for example, in our data the most prominent zones for 61~Cyg~B occur around 1203--1211~MHz and 1265--1272~MHz. The exact boundaries adopted for that target are given in Table~\ref{tab:rfizones}.

\subsection{Application to PSR B0329+54}
\label{sec:psr_analysis}

To validate the performance of the analysis pipeline, we apply the same processing steps to the PSR~B0329+54 dataset described in Section~\ref{sec:psr_obs}. For this test, we restrict the FFA search to a narrower period range bracketing the known pulsar period (0.2144--1.7862~s), with the full configuration summarized in Table~\ref{tab:blipss_parameters_for_pulsar}. We then carry the resulting hits through the multi-beam, polarization, RFI-band, clustering, and working-band stages without modification.

\begin{figure}[ht]
\centering
\begin{minipage}[t]{0.45\textwidth}
    \centering
    \includegraphics[width=\linewidth]{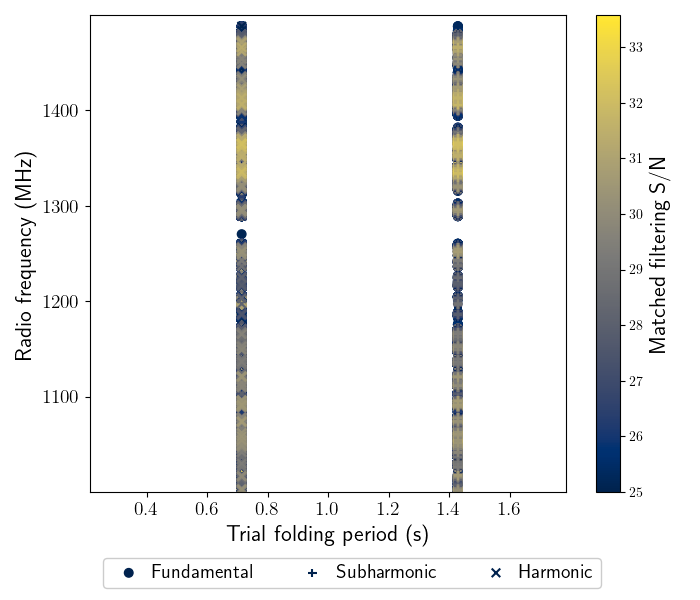}
    \caption{Detection of test pulsar B0329+54 and its harmonics as the two distinct peaks at $\approx 0.71s$ and $1.42 s$.}
    \label{fig:left}
\end{minipage}%
\hfill
\begin{minipage}[t]{0.45\textwidth}
    \centering
    \includegraphics[width=\linewidth]{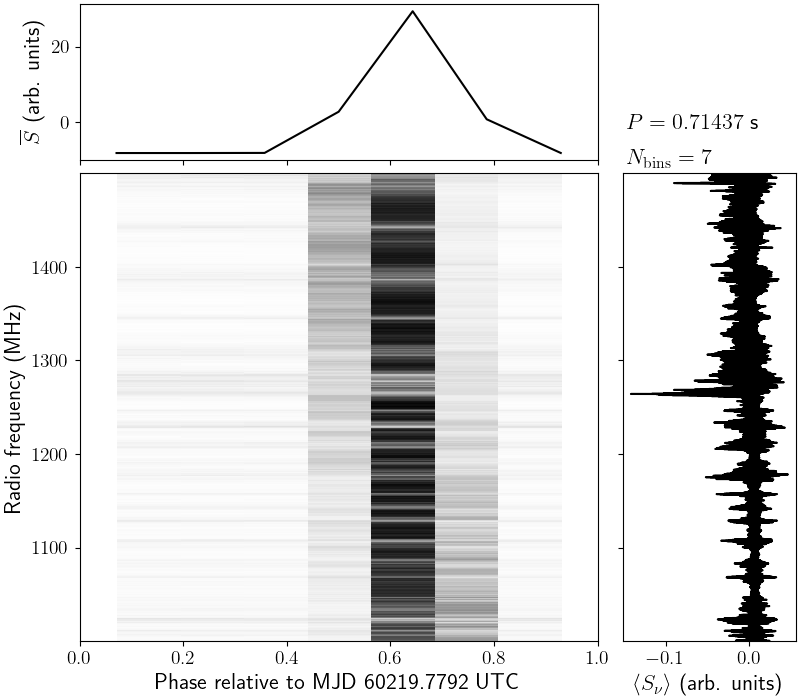}
    \caption{The phase-resolved spectrum of the pulsar B0329+54 based on the period obtained by BLIPSS.The top panel shows the average flux density ($
\overline{S}$) in arbitrary units as a function of the phase relative to MJD 60219.7792 UTC. The bottom panel is a grayscale plot showing the radio frequency (MHz) on the vertical axis and the phase on the horizontal axis.The right panel plots the average flux density ($
\overline{S}$) in arbitrary units as a function of radio frequency.}
    \label{fig:right}
\end{minipage}
\end{figure}

In the resulting cluster-level catalog, the strongest candidate appears at a period $P = 0.7144 \pm 0.0001$\,s and its harmonics, as shown in Figure~\ref{fig:left}, fully consistent with the timing ephemeris from the ATNF catalogue \citep{manchester2005atnf}. Furthermore, as the phase--time diagram in Figure~\ref{fig:right} illustrates, the signal is persistent throughout the observing window and forms a phase-stable vertical stripe, indicating that pulses arrive at essentially the same rotational phase for the entire scan.

Importantly, the pulsar is detected exclusively in the on-axis beam, with Codes consistent with strictly single-beam occupancy, and passes all polarization-consistency and RFI-band cuts. This demonstrates that our pipeline is capable of recovering a known astrophysical periodic signal with the expected period, harmonic content, and beam pattern, and supports the interpretation of our stellar non-detections as robust constraints on periodic technosignatures.

\begin{deluxetable}{ccccc}
\tablecaption{Ad hoc RFI exclusion zones within the 1050--1450\,MHz working band for 61~Cyg~B\label{tab:rfizones}}
\tablehead{
\colhead{Zone ID} &
\colhead{$\nu_{\min}$ (MHz)} &
\colhead{$\nu_{\max}$ (MHz)} &
\colhead{Width (MHz)} &
\colhead{Comments}
}
\startdata
Z1 & 1203.4  & 1210.8  & 7.4  & Broad, persistent multi-beam RFI complex \\
Z2 & 1264.9  & 1272.1  & 7.2  & Broad, persistent multi-beam RFI complex \\
Z3 & 1211.01 & 1211.07 & 0.06 & Very narrow, high-duty-cycle spike \\
Z4 & 1273.0  & 1273.2  & 0.20 & Narrow, strongly variable spike \\
Z5 & 1263.8  & 1264.0  & 0.20 & Narrow, intermittent RFI feature \\
\enddata
\tablecomments{
Frequency intervals correspond to the ''RFI zones'' applied in the analysis of 61~Cyg~B(Section~\ref{sec:analysis}). Any hit whose frequency falls inside one of these intervals is discarded, even if it passes all previous multi-beam, multi-epoch, polarization, and clustering filters.
}
\end{deluxetable}

\begin{table}[ht]
\centering
\begin{tabular}{ccc}
\hline
\hline
Parameter & Value \\
\hline
Minimum period & 0.2144 seconds\\
Maximum period & 1.7862 seconds\\
Number of phase bins & 6 to 17\\
Duty cycle resolution & 0.05\\
Folding period multiplier  & 3\\
Signal-to-noise ratio threshold & 25\\
\hline
\end{tabular}
\caption{Key Parameter Configuration for Pulsar B0329+54 Detection}
\label{tab:blipss_parameters_for_pulsar}
\end{table}

\section{Results and Discussion}
\label{sec:discussion}

\subsection{Results}
\label{sec:result}

The per-channel FFA search produces a very large number of raw hits per target when all beams, polarisations, channels, and trial periods are considered. As expected for ground-based L-band observations, the overwhelming majority of these A0 hits are associated with terrestrial or instrumental signals rather than astrophysical periodicities. The successive RFI-mitigation and consistency filters described in Section~\ref{sec:analysis} reduce this population in a highly structured way, which we quantify in Table~\ref{tab:rfi_flagging_aastex}.

At the raw-output stage A0, the targets exhibit between $\sim 4.7\times 10^{5}$ (Ross~248) and $\sim 2.6\times 10^{6}$ (Ross~128) hits, reflecting both differences in RFI environment and in the number of valid channels per observation. Applying the multi-beam occupancy cut ($\mathrm{RFI1}=0$) reduces these counts by only $\sim 5$--10\% (A1), indicating that strong, simultaneously multi-beam signals form a relatively small subset of the total hit population. By contrast, the long-term channel mask based on $\mathrm{RFI2}$ produces a dramatic reduction for some targets: for 61~Cyg~B, the A2 count drops from $2.28\times 10^{6}$ to $6.83\times 10^{5}$ hits, demonstrating that a substantial fraction of its initial hits originate from a small set of chronically contaminated channels. For the other four targets the A2 counts are reduced by one to two orders of magnitude relative to A0, underscoring the importance of cross-target RFI statistics in identifying persistent bad channels.

The polarization-coincidence filter and the $\mathrm{RFI3}$ flag provide an additional and very strong discrimination step. Requiring that hits be consistent between XX and YY and free of any $\mathrm{RFI1}$, $\mathrm{RFI2}$, or $\mathrm{RFI3}$ flags (A3) reduces the hit counts to tens for Groombridge~34~A/B and Ross~248, and to $\sim 10^{4}$--$10^{5}$ for 61~Cyg~B and Ross~128. The broad frequency-domain mask $\mathrm{RFI\_band}$ (A4) removes a modest fraction of these remaining hits for most targets, but is particularly effective for 61~Cyg~B, where it reduces the hit count from $6.93\times 10^{4}$ to $2.76\times 10^{4}$. This behaviour is consistent with the presence of a few broad RFI complexes in the 61~Cyg~B data (Table~\ref{tab:rfizones}), which are not fully captured by the channel-based $\mathrm{RFI2}$ statistic alone.

\begin{figure*}
    \centering
    \includegraphics[width=0.4\textwidth]{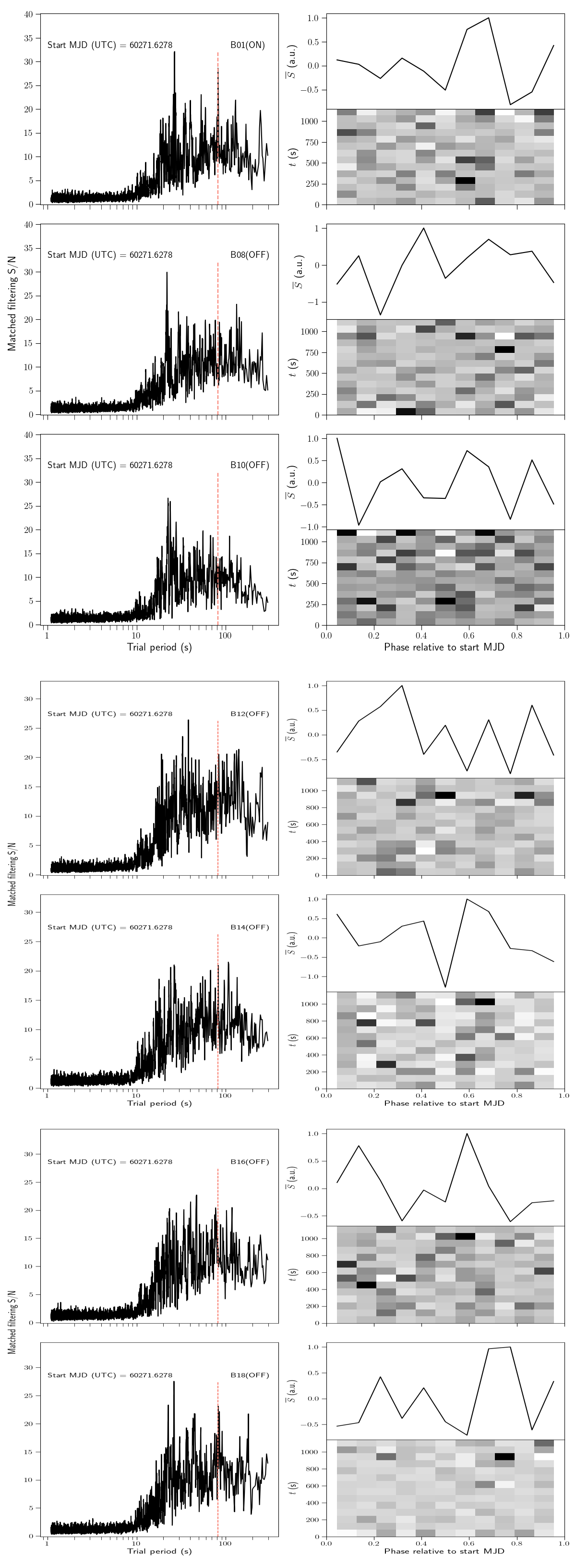}
    \includegraphics[width=0.4\textwidth]{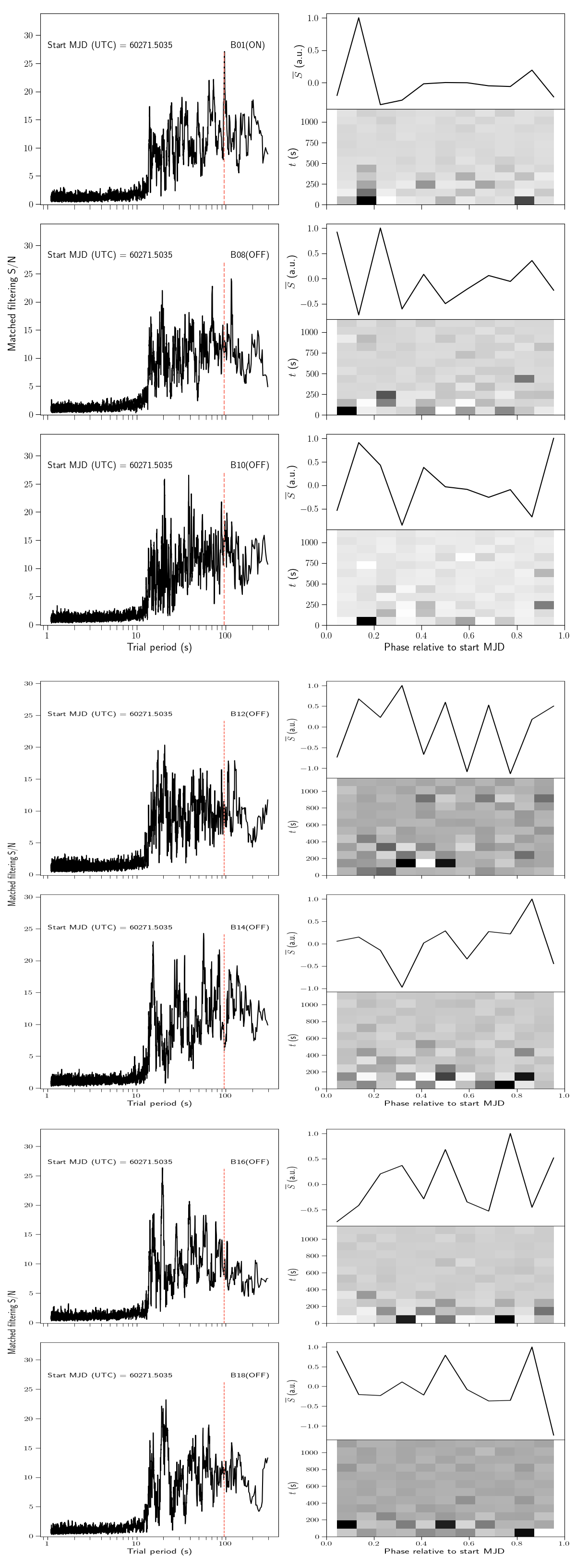}
    \caption{
    Examples of phase--time diagnostic plots for representative high-S/N hits in the final stages of the pipeline.
    Left: A5 hits for Groombridge~34~A, illustrating typical residual RFI morphology in the phase--time domain where a consistent vertical line is absent (short-lived, band-limited bursts or obvious instrumental artefacts).
    Right: A6 hits for 61~Cyg~B after application of the RMS-based RFI exclusion zones (Table~\ref{tab:rfizones}); all exhibit clear RFI-like behaviour rather than stable, strictly periodic technosignatures.
    }
    \label{fig:phasetime_examples}
    \label{fig:phasetime_61cyg}
\end{figure*}

The final automated stage at the hit level, A5, restricts attention to hits that (i) lie within the adopted 1050--1450\,MHz working band and (ii) are detected only in the central beam (Code $=1000000$). After this step, the five targets retain only a handful of hits each: 3 for Groombridge~34~A, 0 for Groombridge~34~B, 1 for Ross~248, 1,282 for 61~Cyg~B, and 0 for Ross~128. For four of the five stars (Groombridge~34~A/B, Ross~248, and Ross~128), the automated pipeline therefore already compresses the hit population down to at most a few A5 hits. For each of these remaining hits we examined the phase--time and phase--frequency diagnostic plots by eye and found that all are clearly attributable to residual RFI (e.g.\ short-lived bursts confined to a narrow set of channels, or features coincident with known instrumental artefacts). Representative examples are shown in Figure~\ref{fig:phasetime_examples}. No plausible astrophysical periodicity remains in the A5 sample for these four targets.

The case of 61~Cyg~B is more challenging. Even after A5, this target retains 1,282 hits that are single-beam, polarization-consistent, and outside the broad $\mathrm{RFI\_band}$ mask. To further quantify persistent RFI in the 1050--1450\,MHz working band for this source, we constructed an RMS spectrum of the combined XX+YY dynamic spectra as a function of frequency and used the median and median absolute deviation (MAD) of the RMS distribution to define robust outliers in frequency space. Channels whose RMS exceeded the band median by more than 5 times of the MAD were flagged and consecutive flagged channels were merged into frequency intervals. Comparison of these intervals with dedicated RFI-monitoring measurements at the FAST site confirmed that they correspond to a small set of frequencies that are persistently contaminated by strong, clearly terrestrial signals. The resulting five ''RFI exclusion zones'' (Z1--Z5) are listed in Table~\ref{tab:rfizones} and include two broad, nearly wall-to-wall complexes (Z1 and Z2) and three narrow, high-duty-cycle or strongly variable spikes (Z3--Z5).

We adopt Z1--Z5 as a final, hard-coded RFI mask for 61~Cyg~B: any hit whose frequency falls within one of these intervals is discarded, even if it passes all previous multi-beam, polarization, and clustering filters. Applying this additional frequency cut to the 1,282 A5 hits defines an A6 working sample of 693 hits for 61~Cyg~B. We then subjected all A6 hits to the same phase--time visual inspection as for the other targets. In every case the diagnostics reveal behaviour characteristic of RFI (e.g.\ bursts confined to short sub-intervals of the observation, structures correlated across many nearby channels, or patterns that repeat across beams in a way incompatible with a point source on the sky), with no candidates displaying the stability and coherence expected of a technosignature. Examples of these A6 phase--time plots for 61~Cyg~B are shown in Figure~\ref{fig:phasetime_61cyg}.

In summary, starting from A0 raw FFA output ranging from $\sim 5\times 10^{5}$ to $\sim 3\times 10^{6}$ hits per target, our multi-stage RFI-mitigation pipeline (A1--A5), combined with a dedicated RMS-based exclusion of local RFI zones for 61~Cyg~B (A6), reduces the candidate population to a small number of cluster-level candidates per target, all of which can be confidently attributed to terrestrial or instrumental origins on the basis of their diagnostic plots. We therefore treat this survey as a non-detection of periodic technosignatures in the searched parameter space and, in the following subsections, convert these non-detections into quantitative limits on the luminosities of putative transmitters.

\begin{deluxetable*}{lccccc} 
    \tablecaption{Hit counts after successive RFI--mitigation steps for each target}
    \label{tab:rfi_flagging_aastex}
    \tablewidth{0pt}
    \tablehead{
        \colhead{Stage} & 
        \colhead{Groombridge 34 A} & 
        \colhead{Groombridge 34 B} & 
        \colhead{Ross 248} & 
        \colhead{61 Cyg B} & 
        \colhead{Ross 128}
    }
    \startdata
    A0 & 553,071 & 742,621 & 471,321 & 2,422,266 & 2,620,968 \\
    A1 & 517,256 & 699,794 & 447,532 & 2,283,293 & 2,301,071 \\
    A2 & 22,531  & 25,907  & 6,979   & 682,511   & 119,730   \\
    A3 & 87      & 278     & 58      & 69,299    & 10,785    \\
    A4 & 14      & 47      & 7       & 27,644    & 2,385     \\
    A5 & 3       & 0       & 1       & 1,282     & 0         \\
    \enddata
    \tablecomments{
    Number of BLIPSS--FFA hits per target after each successive filtering stage. 
    A0: total number of raw FFA hits from both XX and YY polarisations, before any RFI masking. 
    A1: hits remaining after rejecting all candidates with $RFI1 = 1$ (multi-beam occupancy). 
    A2: hits remaining after additionally rejecting channels flagged as long-term contaminants with $RFI2 = 1$. 
    A3: hits remaining after applying the polarization-coincidence filter and removing all hits with $RFI1 = 1$, $RFI2 = 1$, or $RFI3 = 1$. 
    A4: subset of A3 lying outside the broad frequency-domain mask, i.e.\ hits with $RFI\_\mathrm{band} = 0$. 
    A5: final working sample, obtained by selecting A4 hits that fall within the adopted 1050--1450\,MHz working band and are detected only in the central beam.
    }
\end{deluxetable*}

\subsection{Sensitivity to periodic beacons}
\label{sec:results_sensitivity}
 For a pulse train with period $P$, duty cycle $\delta$ (fractional pulse width), and effective bandwidth $\Delta\nu_{\rm eff}$, the minimum detectable phase-averaged flux density at a given frequency is
\begin{equation}
    S_{\min}(P,\delta) = 
    \beta\, \frac{(\mathrm{S/N})_{\rm thr}\, T_{\rm sys}}
    {G\,\sqrt{n_{\rm pol}\, \Delta\nu_{\rm eff}\, T_{\rm obs}}}\,
    \sqrt{\frac{\delta}{1-\delta}},
    \label{eq:sradiometer}
\end{equation}
where $T_{\rm sys}$ is the system temperature, $G$ is the telescope gain, $n_{\rm pol}=2$ is the number of summed polarizations, $T_{\rm obs}$ is the total integration time, $(\mathrm{S/N})_{\rm thr}$ is the adopted FFA detection threshold, and $\beta\approx 1$ accounts for digitization losses.

For the FAST L-band system, we adopt $T_{\rm sys}=25$~K and $G=16$~K~Jy$^{-1}$ for the central beam \citep{jiang2020fastperf}, $T_{\rm obs}=1200$~s, $n_{\rm pol}=2$, and $(\mathrm{S/N})_{\rm thr}=25$ as in Section~\ref{sec:analysis_ffa}. The FFA operates on individual frequency channels, so we take $\Delta\nu_{\rm eff}$ to be comparable to the channel width; to facilitate comparison with previous BLIPSS work at the Green Bank Telescope, which quotes sensitivities for kHz-wide signals \citep{suresh2023blipss}, we express our limits in terms of an effective $\Delta\nu_{\rm eff}=476.8$~kHz. For a representative duty cycle $\delta=0.1$, Equation~\eqref{eq:sradiometer} then yields a minimum detectable phase-averaged flux density of order $S_{\min}\sim 10^{-2}$~Jy at L band, with only a modest dependence on period over the 1.1--300~s range explored.

The duty-cycle dependence enters through the factor $\sqrt{\delta/(1-\delta)}$ in Equation~\eqref{eq:sradiometer}. For narrow pulses with $\delta\ll 1$, our sensitivity improves approximately as $S_{\min}\propto \sqrt{\delta}$, so that more sharply peaked beacons are easier to detect at fixed average power. Conversely, duty cycles approaching 50\% incur a modest penalty. Across the $\delta=0.1$--0.5 range considered here, this dependence is at most a factor of a few and does not qualitatively change our conclusions.

\subsection{EIRP limits for the five targets}
\label{sec:results_eirp}

For each star, we convert the flux-density limits into constraints on the equivalent isotropic radiated power (EIRP) of a putative transmitter located at the distance $d$ of the target. For a beacon whose emission is confined to a bandwidth $\Delta\nu_{\rm eff}$ but is otherwise isotropic, the EIRP limit corresponding to $S_{\min}$ is
\begin{equation}
    \mathrm{EIRP}_{\min} = 4\pi d^{2}\, S_{\min}\, \Delta\nu_{\rm eff}.
    \label{eq:eirp}
\end{equation}
We adopt distances from Gaia~DR3 (Table~\ref{tab:targets}) and the $S_{\min}$ values derived from Equation~\eqref{eq:sradiometer}, evaluated at $\Delta\nu_{\rm eff}=476.8$~kHz and a fiducial duty cycle $\delta=0.1$. The resulting EIRP limits scale as $\mathrm{EIRP}_{\min}\propto d^{2}$ and are therefore very similar for all five targets, which lie between 3.2 and 3.6~pc.

Numerically, for the FAST system parameters described above, the resulting constraints are of order
\begin{equation}
    \mathrm{EIRP}_{\min} \sim \mathrm{few}\times 10^{9}\text{--}10^{10}~\mathrm{W}
\end{equation}
for kHz-wide periodic signals in the 1.05--1.45~GHz band, depending on the exact duty cycle and adopted effective bandwidth. These values are comparable to, or somewhat below, the peak powers of the most powerful human-made radio transmitters and radar systems, and significantly below the EIRP limits typically obtained for similar systems at larger distances \citep{zhang2020fastseti,ma2023dl,suresh2023blipss}. The detailed star-by-star limits as a function of duty cycle and assumed bandwidth are summarized in Table~\ref{tab:limits}, which serves as the basis for the population-level discussion below.

We emphasize that Equation~\eqref{eq:eirp} assumes isotropic emission. If extraterrestrial transmitters preferentially beam their radiation toward Earth, the intrinsic transmitter power could be lower than our EIRP limits by the beaming factor. Conversely, if beacons are intermittingly active (e.g.\ only during a fraction of our observing window), our effective sensitivity to their time-averaged power is reduced. In this study we do not attempt to correct for such duty-cycle effects on the transmitter side; instead, we treat our EIRP limits as constraints on the product of transmitter power and beaming toward Earth.
\begin{deluxetable}{lcc}
\tablecaption{Indicative upper limits on isotropic-equivalent EIRP for periodic transmitters\label{tab:limits}}
\tablehead{
\colhead{Target} &
\colhead{$d$ (pc)} &
\colhead{$L_{\mathrm{EIRP,min}}$ ($10^{9}$\,W)}
}
\startdata
Groombridge~34 A & 3.5625 & 8.8 \\
Groombridge~34 B & 3.5638 & 8.8 \\
Ross~248         & 3.4947 & 8.5 \\
61~Cyg~B         & 3.1598 & 7.0 \\
Ross~128         & 3.3731 & 7.9 \\
\enddata
\tablecomments{
Values are formal upper limits on the isotropic-equivalent radiated power of a periodically modulated, narrowband transmitter at each target, computed using the radiometer equation with a detection threshold of $(\mathrm{S/N})_{\rm thr}=25$ and a duty cycle of 0.1 of a periodically modulated, narrowband transmitter at each target, assuming the FAST+BLIPSS sensitivity and search setup described in Sections~\ref{sec:analysis}. For fixed instrumental assumptions the limits scale as $L_{\mathrm{EIRP,min}}\propto d^{2}$.
}
\end{deluxetable}

\subsection{Context within technosignature searches}
\label{sec:results_context}

The non-detections reported here add a new class of constraints to the technosignature literature. Previous FAST SETI observations have focused primarily on narrowband continuous-wave signals toward individual nearby stars and exoplanet systems, typically quoting EIRP limits for 1--3~Hz-wide signals \citep{zhang2020fastseti}. Other large programs, such as Breakthrough Listen with the Green Bank Telescope and Parkes/Murriyang, have extended these narrowband searches to thousands of nearby stars over wide frequency ranges \citep{ma2023dl,wright2022techno}. The BLIPSS search at 4--8~GHz, by contrast, pioneered per-channel FFA-based searches for periodic beacons toward the Galactic Center \citep{suresh2023blipss}.

Our work complements these efforts in three ways. First, it demonstrates that FFA-based searches for periodic technosignatures can be successfully ported from GBT to FAST, exploiting the latter's multi-beam architecture and sensitivity at L band. The end-to-end validation on PSR~B0329+54 (Section~\ref{sec:psr_analysis}) shows that the combination of BLIPSS and our multi-layer RFI pipeline recovers known astrophysical periodicities without being overwhelmed by terrestrial interference. Second, this is, to our knowledge, the first application of an FFA-based technosignature search to some of the nearest known stellar systems with strong exoplanet and habitability interest: Groombridge~34, Ross~248, 61~Cygni~B, and Ross~128. The resulting EIRP limits in the $\sim 10^{9}$--$10^{10}$~W range are among the most stringent constraints to date on kHz-wide periodic beacons in the L band from any nearby exoplanet system.

Third, by deliberately combining a heterogeneous sample --- including an exoplanet host (Ross~128), a multi-planet M dwarf (Groombridge~34~A), a K-dwarf benchmark (61~Cygni~B), and a planet-free control (Ross~248) --- our survey provides a template for future FAST programs that aim to link technosignature limits to stellar and planetary demographics. 

\begin{figure*}
    \centering
    \includegraphics[width=0.66\textwidth]{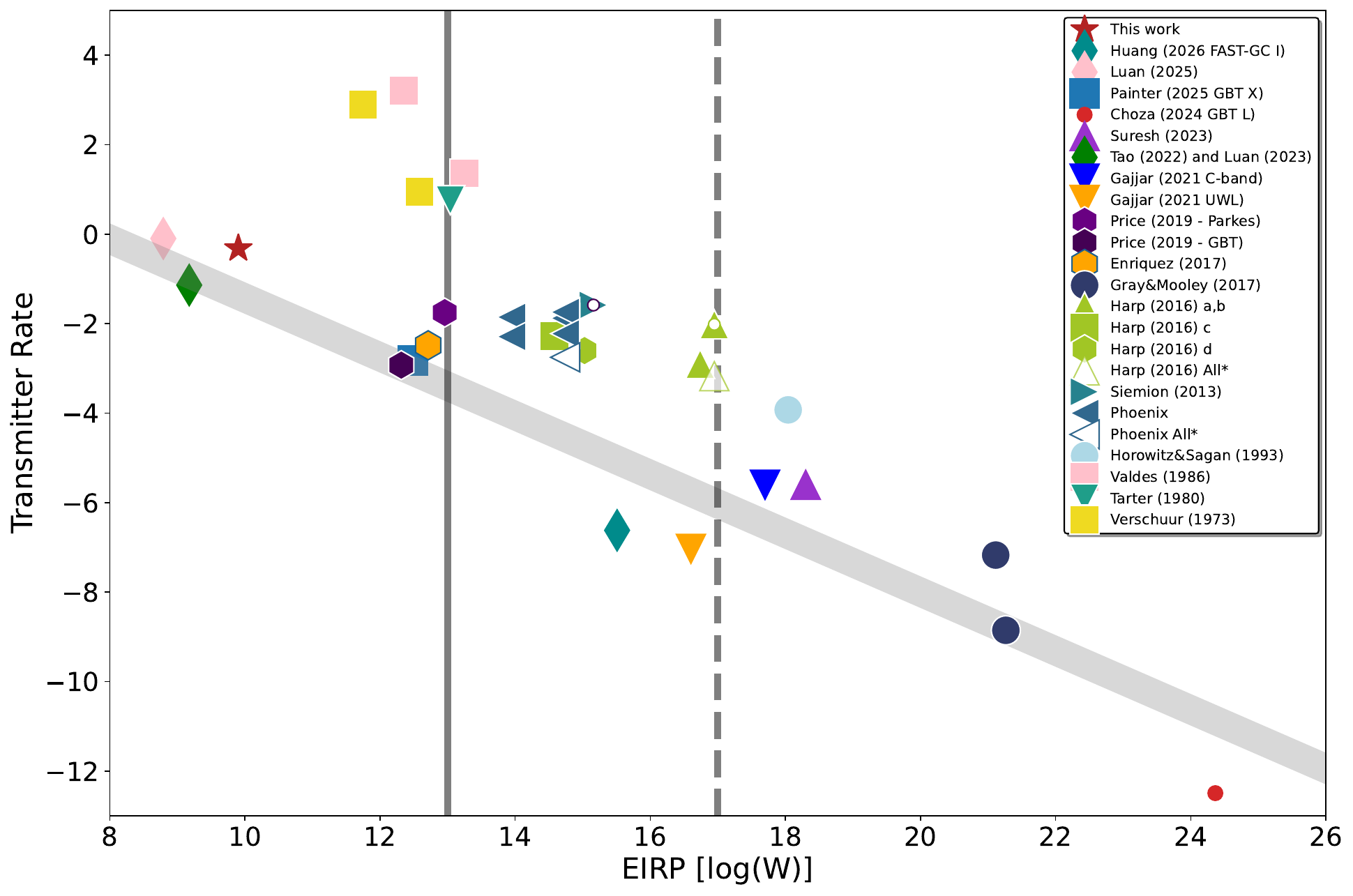}
    \caption{
   Comparison of this work (dark-red star) with previous radio technosignature surveys in the plane of EIRP sensitivity and an effective transmitter-rate figure of merit.
    The horizontal axis shows the characteristic upper limit on the EIRP of detectable transmitters (lower values indicate higher sensitivity), while the vertical axis encodes how strongly each survey constrains the space density or occurrence rate of powerful transmitters (lower values indicate tighter population limits).
    Our FAST periodic search occupies the upper-left region of the diagram, corresponding to very high EIRP sensitivity but relatively weak constraints on transmitter rates because only a small number of nearby targets were observed.
    }
    \label{fig:transmitter_rate_eirp}
\end{figure*}

A quantitative comparison of this work with previous radio technosignature searches is shown in Figure~\ref{fig:transmitter_rate_eirp}, which places a number of surveys in the plane of EIRP sensitivity versus an effective ``transmitter rate'' figure of merit. The horizontal axis gives the characteristic upper limit on the EIRP of detectable transmitters, so points further to the left correspond to deeper, more sensitive searches. The vertical axis is constructed such that lower values indicate stronger constraints on the occurrence rate of powerful transmitters (typically because many targets or large areas of sky were surveyed), whereas higher values correspond to weaker constraints. Our FAST periodic search (dark-red diamond) lies in the upper-left of this diagram: it is among the most sensitive surveys in EIRP, capable of detecting kHz-wide periodic beacons with powers below those reached by most previous programs, but it constrains transmitter rates only weakly because it targets just five nearby stars. In this sense, our survey complements large-sample campaigns that provide stronger population-level limits at higher EIRP thresholds, and together these programs map out a broad region of the EIRP--transmitter-rate trade space.

At the same time, several caveats are worth emphasizing. Our search is sensitive only to strictly periodic signals within the 1.1--300~s period range, and only if their emission is confined to roughly kHz-wide channels near 1.05--1.45~GHz. Transmitters employing broader-band modulation, highly intermittent emission, or strongly frequency-drifting signals could evade detection in the present analysis. Likewise, beacons operating outside FAST L band, or using exotic dispersion profiles that move power outside a single channel, would not be captured. Future work could address these limitations by extending FFA-based searches to a wider period range, incorporating explicit searches for non-zero dispersion, and combining periodic and narrowband strategies within a unified technosignature framework.

\section{Conclusions and Future Prospects}
\label{sec:conclusion}

We have presented a search for periodic technosignatures toward five of the nearest stellar systems observable with FAST: Groombridge~34~A, Groombridge~34~B, Ross~248, 61~Cygni~B, and Ross~128. Using the FAST 19-beam L-band receiver, we obtained 1200\,s tracking observations for each target and analysed the resulting 1.05--1.45\,GHz dynamic spectra with a per-channel Fast Folding Algorithm (FFA) search for strictly periodic signals with periods between 1.1 and 300\,s. The targets are all K--M dwarfs within $\sim 4$\,pc, selected for their proximity and strong exoplanet or habitability interest, including the temperate Earth-mass planet host Ross~128~b \citep{bonfils2018ross128b} and the nearby multi-planet system Groombridge~34~A \citep{howard2014gj15ab,pinamonti2018gj15a}.

Our results complement previous FAST and Green Bank technosignature efforts that have focused primarily on continuous narrowband signals \citep[e.g.,][]{zhang2020fastseti,huang2023solution} or periodic beacons toward the Galactic Center \citep{suresh2023blipss}. In particular, they demonstrate that FFA-based searches for periodic technosignatures can be robustly implemented on FAST, leveraging its multi-beam architecture and high sensitivity at L band. They also provide a template for future programs that aim to connect technosignature limits with the demographics of nearby stellar and planetary systems, by combining exoplanet hosts, K-dwarf benchmarks, and planet-free controls in a single, homogeneous survey.

Several limitations of the present study point naturally to future work. First, our search is restricted to strictly periodic signals with periods between 1.1 and 300\,s and approximate duty cycles between 0.1 and 0.5, confined to $\sim$kHz-wide channels at 1.05--1.45\,GHz. Beacons with much longer or shorter periods, strongly time-variable duty cycles, broader-band modulation, or significant frequency drift could evade our analysis. Second, we have not explicitly searched for non-zero dispersion measures; while dispersion is negligible for nearby systems in the interstellar medium, more exotic propagation effects cannot be ruled out. Third, each target currently has only a single 1200\,s epoch, which limits our sensitivity to intermittently active transmitters.

Looking ahead, the most promising extensions of this work include: (i) multi-epoch observations of the same targets to probe intermittent or temporally clustered transmissions; (ii) application of the pipeline to larger samples of nearby stars with well-characterized planetary systems, enabling statistical constraints on the fraction of habitable-zone planets hosting periodic beacons; (iii) integration of FFA-based searches with simultaneous narrowband and transient technosignature pipelines on FAST; and (iv) expansion to other frequency bands accessible to FAST and complementary facilities. As the broader technosignature field continues to grow \citep{wright2022techno}, such coordinated, multi-modal surveys will be essential for translating diverse non-detections into quantitative statements about the prevalence, power, and signalling strategies of potential extraterrestrial technologies.

\section*{Acknowledgment}

This work was supported by the National Key R$\&$D Program of China, No.2024YFA1611804 and the China Manned Space Program with grant No. CMS-CSST2025-A01 and Shandong Provincial Natural Science Foundation (ZR2024QA180) and Scientific Research Fund of Dezhou University (4022504019).This work made use of the data from FAST (Five-hundred-meter Aperture Spherical radio Telescope). FAST is a Chinese national mega-science facility, operated by National Astronomical Observatories, Chinese Academy of Sciences.

\bibliography{sample701}{}

@article{Kopparapu_2013,
   title={HABITABLE ZONES AROUND MAIN-SEQUENCE STARS: NEW ESTIMATES},
   volume={765},
   ISSN={1538-4357},
   url={http://dx.doi.org/10.1088/0004-637X/765/2/131},
   DOI={10.1088/0004-637x/765/2/131},
   number={2},
   journal={The Astrophysical Journal},
   publisher={American Astronomical Society},
   author={Kopparapu, Ravi Kumar and Ramirez, Ramses and Kasting, James F. and Eymet, Vincent and Robinson, Tyler D. and Mahadevan, Suvrath and Terrien, Ryan C. and Domagal-Goldman, Shawn and Meadows, Victoria and Deshpande, Rohit},
   year={2013},
   month=feb, pages={131} }

@article{Valle_2014,
   title={Evolution of the habitable zone of low-mass stars: Detailed stellar models and analytical relationships for different masses and chemical compositions⋆},
   volume={567},
   ISSN={1432-0746},
   url={http://dx.doi.org/10.1051/0004-6361/201323350},
   DOI={10.1051/0004-6361/201323350},
   journal={Astronomy \& Astrophysics},
   publisher={EDP Sciences},
   author={Valle, G. and Dell’Omodarme, M. and Prada Moroni, P. G. and Degl’Innocenti, S.},
   year={2014},
   month=jul, pages={A133} }

@article{Sheikh_2021,
   title={Analysis of the Breakthrough Listen signal of interest blc1 with a technosignature verification framework},
   volume={5},
   ISSN={2397-3366},
   url={http://dx.doi.org/10.1038/s41550-021-01508-8},
   DOI={10.1038/s41550-021-01508-8},
   number={11},
   journal={Nature Astronomy},
   publisher={Springer Science and Business Media LLC},
   author={Sheikh, Sofia Z. and Smith, Shane and Price, Danny C. and DeBoer, David and Lacki, Brian C. and Czech, Daniel J. and Croft, Steve and Gajjar, Vishal and Isaacson, Howard and Lebofsky, Matt and MacMahon, David H. E. and Ng, Cherry and Perez, Karen I. and Siemion, Andrew P. V. and Webb, Claire Isabel and Zic, Andrew and Drew, Jamie and Worden, S. Pete},
   year={2021},
   month=oct, pages={1153–1162} }

@article{Tao_2023,
   title={The Most Sensitive SETI Observations Toward Barnard’s Star with FAST},
   volume={166},
   ISSN={1538-3881},
   url={http://dx.doi.org/10.3847/1538-3881/acfc1e},
   DOI={10.3847/1538-3881/acfc1e},
   number={5},
   journal={The Astronomical Journal},
   publisher={American Astronomical Society},
   author={Tao, Zhen-Zhao and Huang, Bo-Lun and Luan, Xiao-Hang and Li, Jian-Kang and Zhao, Hai-Chen and Wang, Hong-Feng and Zhang 张, Tong-Jie 同杰},
   year={2023},
   month=oct, pages={190} }

@article{cocconi1959,
  author       = {Cocconi, Giuseppe and Morrison, Philip},
  title        = {Searching for Interstellar Communications},
  journal      = {Nature},
  volume       = {184},
  number       = {4690},
  pages        = {844--846},
  year         = {1959},
  doi          = {10.1038/184844a0}
}

@article{tarter2001seti,
  author       = {Tarter, Jill C.},
  title        = {The Search for Extraterrestrial Intelligence (SETI)},
  journal      = {Annual Review of Astronomy and Astrophysics},
  volume       = {39},
  pages        = {511--548},
  year         = {2001},
  doi          = {10.1146/annurev.astro.39.1.511}
}

@article{wright2022techno,
  author       = {Wright, Jason T. and Haqq-Misra, Jacob and Frank, Adam and Kopparapu, Ravi and Lingam, Manasvi and Sheikh, Sofia Z.},
  title        = {The Case for Technosignatures: Why They May Be Abundant, Long-lived, Highly Detectable, and Unambiguous},
  journal      = {Astrophysical Journal Letters},
  volume       = {927},
  number       = {2},
  pages        = {L30},
  year         = {2022},
  doi          = {10.3847/2041-8213/ac5824}
}

@article{suresh2023blipss,
  author       = {Suresh, Akshay and Gajjar, Vishal and Nagarajan, Pranav and Sheikh, Sofia Z. and Siemion, Andrew P. V. and Lebofsky, Matt and MacMahon, David H. E. and Price, Danny C. and Croft, Steve},
  title        = {A 4--8 GHz Galactic Center Search for Periodic Technosignatures},
  journal      = {The Astronomical Journal},
  volume       = {165},
  number       = {6},
  pages        = {255},
  year         = {2023},
  doi          = {10.3847/1538-3881/acccf0},
  eprint       = {2305.18527},
  archivePrefix= {arXiv}
}

@article{jiang2020fastperf,
  author       = {Jiang, Peng and Tang, Ning-Yu and Hou, Li-Gang and Liu, Meng-Ting and Kr{\v{c}}o, Marko and Qian, Lei and Sun, Jing-Hai and Ching, Tao-Chung and Liu, Bin and Duan, Yan and Yue, You-Ling and Gan, Heng-Qian and Yao, Rui and Li, Hui and Pan, Gao-Feng and Yu, Dong-Jun and Liu, Hong-Fei and Li, Di and Peng, Bo and Yan, Jun and FAST Collaboration},
  title        = {The Fundamental Performance of {FAST} with 19-beam Receiver at {L} Band},
  journal      = {Research in Astronomy and Astrophysics},
  volume       = {20},
  number       = {5},
  pages        = {064},
  year         = {2020},
  doi          = {10.1088/1674-4527/20/5/64},
  eprint       = {2002.01786},
  archivePrefix= {arXiv}
}

@article{zhang2020fastseti,
  author       = {Zhang, Zhi-Song and Werthimer, Dan and Zhang, Tong-Jie and Cobb, Jeff and Korpela, Eric and Anderson, David and Gajjar, Vishal and Lee, Ryan and Li, Shi-Yu and Pei, Xin and Zhang, Xin-Xin and Huang, Shi-Jie and Wang, Pei and Zhu, Yan and Duan, Ran and Zhang, Hai-Yan and Jin, Cheng-Jin and Zhu, Li-Chun and Li, Di},
  title        = {First {SETI} Observations with {China}'s Five-hundred-meter Aperture Spherical Radio Telescope ({FAST})},
  journal      = {The Astrophysical Journal},
  volume       = {891},
  number       = {2},
  pages        = {174},
  year         = {2020},
  doi          = {10.3847/1538-4357/ab7376},
  eprint       = {2002.02130},
  archivePrefix= {arXiv}
}

@article{bonfils2018ross128b,
  author       = {Bonfils, X. and Astudillo-Defru, N. and Wittrock, J. and others},
  title        = {A Temperate Exo-Earth around a Quiet {M} Dwarf at 3.4 pc},
  journal      = {Astronomy \& Astrophysics},
  volume       = {613},
  pages        = {A25},
  year         = {2018},
  doi          = {10.1051/0004-6361/201731973}
}

@article{turnbull2003habcat,
  author       = {Turnbull, Margaret C. and Tarter, Jill C.},
  title        = {Target Selection for {SETI}. I. A Catalog of Nearby Habitable Stellar Systems},
  journal      = {The Astrophysical Journal Supplement Series},
  volume       = {145},
  number       = {1},
  pages        = {181--198},
  year         = {2003},
  doi          = {10.1086/345779},
  eprint       = {astro-ph/0210675},
  archivePrefix= {arXiv}
}

@phdthesis{ransom2001,
  author       = {Ransom, Scott M.},
  title        = {New Search Techniques for Binary Pulsars},
  school       = {Harvard University},
  year         = {2001}
}

@article{ma2023dl,
  author       = {Ma, Peter Xiangyuan and Ng, Cherry and Rizk, Leandro and Croft, Steve and Siemion, Andrew P. V. and Brzycki, Bryan and Czech, Daniel and Drew, Jamie and Gajjar, Vishal and Hoang, John and Isaacson, Howard and Lebofsky, Matt and MacMahon, David H. E. and de Pater, Imke and Price, Danny C. and Sheikh, Sofia Z. and Worden, S. Pete},
  title        = {A Deep-learning Search for Technosignatures of 820 Nearby Stars},
  journal      = {Publications of the Astronomical Society of the Pacific},
  volume       = {135},
  number       = {1049},
  pages        = {014501},
  year         = {2023},
  doi          = {10.1088/1538-3873/acaf10},
  eprint       = {2301.12670},
  archivePrefix= {arXiv}
}

@article{howard2014gj15ab,
  author       = {Howard, A. W. and Marcy, G. W. and Fischer, D. A. and Isaacson, H. and Muirhead, P. S. and others},
  year         = {2014},
  title        = {The {NASA}-{UC}-{UH} {Eta-Earth} Program. {IV}. {A} Low-mass Planet Orbiting an {M} Dwarf 3.6 pc from {E}arth},
  journal      = {The Astrophysical Journal},
  volume       = {794},
  number       = {1},
  pages        = {51},
  doi          = {10.1088/0004-637X/794/1/51}
}

@article{pinamonti2018gj15a,
  author       = {Pinamonti, M. and Damasso, M. and Marzari, F. and Sozzetti, A. and Maldonado, J. and others},
  year         = {2018},
  title        = {The {HADES} {RV} Programme with {HARPS-N} at {TNG}. {VIII}. {Gl}~15{A}: A multiple wide planetary system sculpted by binary interaction},
  journal      = {Astronomy \& Astrophysics},
  volume       = {617},
  pages        = {A104},
  doi          = {10.1051/0004-6361/201732535}
}

@article{gaia2023dr3,
  author       = {{Gaia Collaboration} and Vallenari, A. and Brown, A. G. A. and Prusti, T. and de Bruijne, J. H. J. and others},
  year         = {2023},
  title        = {{Gaia} Data Release 3: Summary of the content and survey properties},
  journal      = {Astronomy \& Astrophysics},
  volume       = {674},
  pages        = {A1},
  doi          = {10.1051/0004-6361/202243940}
}

@article{qian2020fastreview,
  author       = {Qian, L. and Pan, Z. and Li, D. and others},
  year         = {2020},
  title        = {Review of {FAST}: Its scientific achievements and prospects},
  journal      = {The Innovation},
  volume       = {1},
  number       = {3},
  pages        = {100053},
  doi          = {10.1016/j.xinn.2020.100053}
}

@article{gautier2007ross248,
  author       = {Gautier, T. N. and Rieke, G. H. and Stansberry, J. and others},
  year         = {2007},
  title        = {Far-infrared properties of nearby {M} dwarfs},
  journal      = {The Astrophysical Journal},
  volume       = {667},
  pages        = {527--543},
  doi          = {10.1086/520917}
}

@article{manchester2005atnf,
  author       = {Manchester, R. N. and Hobbs, G. B. and Teoh, A. and Hobbs, M.},
  year         = {2005},
  title        = {The {Australia Telescope National Facility} pulsar catalogue},
  journal      = {The Astronomical Journal},
  volume       = {129},
  number       = {4},
  pages        = {1993--2006},
  doi          = {10.1086/428488}
}

@article{huang2023solution,
  title={A solution to continuous RFI in narrowband radio SETI with FAST: The MultiBeam Point-source Scanning strategy},
  author={Huang, Bo-Lun and Tao, Zhen-Zhao and Zhang, Tong-Jie},
  journal={The Astronomical Journal},
  volume={166},
  number={6},
  pages={245},
  year={2023},
  doi={10.3847/1538-3881/ad06b1},
  publisher={IOP Publishing}
}

@article{luan2023multibeam,
  title={Multibeam Blind Search of Targeted SETI Observations toward 33 Exoplanet Systems with FAST},
  author={Luan, Xiao-Hang and Tao, Zhen-Zhao and Zhao, Hai-Chen and Huang, Bo-Lun and Li, Shi-Yu and Liu, Cong and Wang, Hong-Feng and Liu, Wen-Fei and Zhang, Tong-Jie and Gajjar, Vishal and others},
  journal={The Astronomical Journal},
  volume={165},
  number={3},
  pages={132},
  year={2023},
  doi={10.3847/1538-3881/acb706},
  publisher={IOP Publishing}
}

@article{huang2025fast,
  title={The FAST-SETI Milky Way Globular Cluster Survey. I. A Pilot Multibeam On-the-fly Search of Five Globular Clusters at the L Band},
  author={Huang, Bo-Lun and Tao, Zhen-Zhao and Zhang, Tong-Jie and Gajjar, Vishal},
  journal={The Astronomical Journal},
  volume={171},
  number={1},
  pages={51},
  year={2025},
  doi={10.3847/1538-3881/ae2470},
  publisher={IOP Publishing}
}

@article{tao2022sensitive,
  title={Sensitive multibeam targeted SETI observations toward 33 exoplanet systems with FAST},
  author={Tao, Zhen-Zhao and Zhao, Hai-Chen and Zhang, Tong-Jie and Gajjar, Vishal and Zhu, Yan and Yue, You-Ling and Zhang, Hai-Yan and Liu, Wen-Fei and Li, Shi-Yu and Zhang, Jian-Chen and others},
  journal={The Astronomical Journal},
  volume={164},
  number={4},
  pages={160},
  year={2022},
  publisher={IOP Publishing}
}
\bibliographystyle{aasjournalv7}



\end{document}